\documentstyle[preprint,tighten,floats,pra,epsfig,aps]{revtex}
\begin{document}

\draft

\title{Is the concept of quantum probability consistent with
Lorentz covariance?}

\author{Y. S. Kim}
\address{Department of Physics, University of Maryland, \\
College Park, Maryland 20742, U.S.A. \\E-mail: yskim@physics.umd.edu}

\author{Marilyn E. Noz}
\address{Department of Radiology, New York University, \\
New York, New York 10016, U.S.A. \\E-mail: noz@nucmed.med.nyu.edu}

\maketitle

\begin{abstract}
Lorentz-covariant harmonic oscillator wave functions are constructed
from the Lorentz-invariant oscillator differential equation of Feynman,
Kislinger, and Ravndal for a two-body bound state.  The wave
functions are not invariant but covariant.  As the differential
equation contains the time-separation variable, the wave functions
contain the same time-separation variable which does not exist
in Schr\"odinger wave functions.  This time-separation variable can
be shown to belong to Feynman's rest of the universe, and can thus
be eliminated from the density matrix.  The covariant probability
interpretation is given.  This oscillator formalism explains Feynman's
decoherence mechanism which is exhibited in Feynman's parton picture.
\end{abstract}

\section{Introduction}\label{intro}

The concept of localized probability distribution is the backbone of
the present formulation of quantum mechanics.  Of course, we would
like to have more deterministic form of dynamics, and efforts have
been and are still being made along this direction.  One of the most
serious problems with this probabilistic interpretation is whether
this concept of probability is consistent with Lorentz covariance.

In a given Lorentz frame, we know how to do quantum mechanics with
a localized probability distribution.  How would this distribution
look to an observer in a different Lorentz frame?

\begin{itemize}
\item  Would the probability distribution appear the same to this
      observer?

\item  If different, how is the probability distribution distorted?

\item  Is the total probability conserved?

\item  The Lorentz boost mixes up the spatial coordinate with the
       time variable.  What role does the time-separation variable
       play in defining the boundary condition for localization and
       the probability distribution?
\end{itemize}

We can answer some or all of the above questions only if we construct
covariant wave functions, namely wave functions which can be
Lorentz-boosted.  It is easy to construct these wave functions if we
know the answer.  In the initial development of quantum mechanics,
the harmonic oscillator played the pivotal role.  Thus, it is clear
to us that if there is a wave function which can be Lorentz-boosted,
this has to be the harmonic oscillator wave function.  Until we
construct wave functions which can be Lorentz-transformed, we cannot
say that quantum mechanics is consistent with relativity.  Indeed,
we should examine this problem before attempting to invent more
definitive quantum mechanics.

Since the hadron, in the quark model, is a bound-state of
quarks~\cite{gell64}, Feynman, Kislinger, and Ravndal, in
1971~\cite{fkr71}, raised the following question in connection with
the quark model for hadrons: the hadronic spectrum can indeed be
explained in terms of the degeneracy of three-dimensional harmonic
oscillator wave functions in the hadronic rest frame; however, what
happens when the hadron moves? Indeed, Feynman
{\it et al.} wrote a Lorentz-invariant differential equation whose
solutions can become non-relativistic wave functions if the
time-separation variable can be ignored.

This Lorentz-invariant differential equation is a four-dimensional
partial differential equation, with many different solutions depending
on separation of variables and boundary conditions.
There is a set of normalizable solutions which can serve as
a representation space for Wigner's little group for massive
particles~\cite{wig39,knp86}.  We can start with this set of solutions
and give physical interpretations, especially to the time-separation
variable.

We solve this problem using the entropy coming from Feynman's
rest of the universe.  Feynman was interested in the concept of
entropy coming from measurement processes which are less than
complete~\cite{fey72}.  This subject was of course originated by
von Neumann in his classic book on mathematical foundations of
quantum mechanics~\cite{neu32}, but Feynman in his book gives
a very concise explanation of the variable which cannot be observed
and thus belongs to the rest of the universe.  We treat the
time-separation variable as a variable in the rest of the universe.
In this way, we can give a probability interpretation to the
covariant harmonic oscillator wave functions.

Finally, is the covariance of the oscillator wave function consistent
with what we observe in the real world.  Here again, Feynman plays
the key role.  While the quark model can be fit into the oscillator
scheme in the hadronic rest frame, Feynman in 1969 came up with the
idea of partons~\cite{fey69}.  According to Feynman's parton model,
the hadron consists of an infinite number of partons when it moves
with a velocity close to that of light.  Quarks and partons are
believed to be the same particles, but their properties are totally
different.  While the quarks inside the hadron interact coherently
with external signals, partons interact incoherently.  If the partons
are Lorentz-boosted quarks, does the Lorentz boost destroy the
coherence?  We address this question in this paper.

In Sec.~\ref{covham}, we construct the covariant harmonic oscillator
wave functions.  These wave functions can be Lorentz-boosted, but
they depend on the time-separation variable.  In Sec.~\ref{prob},
we give the probability interpretation to the oscillator wave
functions after taking care of the time-separation variable according
to Feynman's rest of the universe.  It is shown in Sec.~\ref{par}.
the quark and parton models are two different manifestation of the
same covariant entity.  The most controversial aspect of Feynman's
parton picture is that the partons interact incoherently with
external signals. In Sec.~\ref{measure}, we show this happens.

%----------------------------------------------------------------------
\begin{figure}%[thb]
\centerline{\epsfig{figure=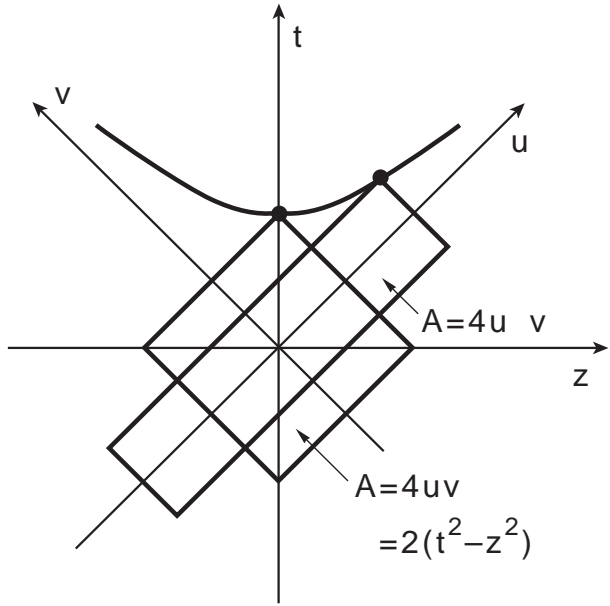,angle=0,width=80mm}}
\vspace{5mm}
\caption{Lorentz boost in the light-cone coordinate
system.}\label{licone}
\end{figure}
%----------------------------------------------------------------------

\section{Can harmonic oscillators be made covariant?}\label{covham}

Quantum field theory has been quite successful in terms of
perturbation techniques in quantum electrodynamics.  However, this
formalism is based on the S matrix for scattering problems and useful
only for physical processes where a set of free particles becomes
another set of free particles after interaction.  Quantum field theory
does not address the question of localized probability distributions
and their covariance under Lorentz transformations.
The Schr\"odinger quantum mechanics of the hydrogen atom deals with
localized probability distribution.  Indeed, the localization condition
leads to the discrete energy spectrum.  Here, the uncertainty relation
is stated in terms of the spatial separation between the proton and
the electron.  If we believe in Lorentz covariance, there must also
be a time-separation between the two constituent particles.

Before 1964~\cite{gell64}, the hydrogen atom was used for
illustrating bound states.  These days, we use hadrons which are
bound states of quarks.  Let us use the simplest hadron consisting of
two quarks bound together with an attractive force, and consider their
space-time positions $x_{a}$ and $x_{b}$, and use the variables
\begin{equation}
X = (x_{a} + x_{b})/2 , \qquad x = (x_{a} - x_{b})/2\sqrt{2} .
\end{equation}
The four-vector $X$ specifies where the hadron is located in space and
time, while the variable $x$ measures the space-time separation
between the quarks.  According to Einstein, this space-time separation
contains a time-like component which actively participates as can be
seen from
\begin{equation}\label{boostm}
\pmatrix{z' \cr t'} = \pmatrix{\cosh \eta & \sinh \eta \cr
\sinh \eta & \cosh \eta } \pmatrix{z \cr t} ,
\end{equation}
when the hadron is boosted along the $z$ direction.
In terms of the light-cone variables defined as~\cite{dir49}
\begin{equation}
u = (z + t)/\sqrt{2} , \qquad v = (z - t)/\sqrt{2} ,
\end{equation}
the boost transformation of Eq.(\ref{boostm}) takes the form
\begin{equation}\label{lorensq}
u' = e^{\eta } u , \qquad v' = e^{-\eta } v .
\end{equation}
The $u$ variable becomes expanded while the $v$ variable becomes
contracted.

Does this time-separation variable exist when the hadron is at rest?
Yes, according to Einstein.  In the present form of quantum mechanics,
we pretend not to know anything about this variable.  Indeed, this
variable belongs to Feynman's rest of the universe.  In this report,
we shall see the role of this time-separation variable in the
decoherence mechanism.

\begin{figure}%[thb]
\centerline{\epsfig{figure=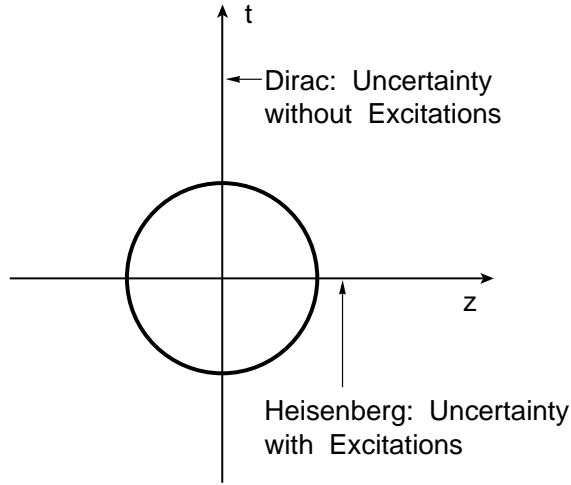,angle=0,width=80mm}}
\vspace{5mm}
\caption{Space-time picture of quantum mechanics.  There
are quantum excitations along the space-like longitudinal direction,
but there are no excitations along the time-like direction.  The
time-energy relation is a c-number uncertainty relation.}\label{quantum}
\end{figure}

Also in the present form of quantum mechanics, there is an uncertainty
relation between the time and energy variables.  However, there are
no known time-like excitations.  Unlike Heisenberg's
uncertainty relation applicable to position and momentum, the time and
energy separation variables are c-numbers, and we are not allowed to
write down the commutation relation between them.  Indeed, the
time-energy uncertainty relation is a c-number uncertainty
relation~\cite{dir27}, as is illustrated in Fig.~\ref{quantum}

%----------------------------------------------------------------------
%-----------------------------------------------------------------------

How does this space-time asymmetry fit into the world of
covariance~\cite{kn73}.  This question was studied in depth by the
present authors in the past.  The answer is that Wigner's $O(3)$-like
little group is not a Lorentz-invariant symmetry, but is a covariant
symmetry~\cite{wig39}.  It has been shown that the time-energy
uncertainty applicable to the time-separation variable fits perfectly
into the $O(3)$-like symmetry of massive relativistic
particles~\cite{knp86}.

The c-number time-energy uncertainty relation allows us to write down
a time distribution function without excitations~\cite{knp86}.
If we use Gaussian forms for both space and time distributions, we
can start with the expression
\begin{equation}\label{ground}
\left({1 \over \pi} \right)^{1/2}
\exp{\left\{-{1 \over 2}\left(z^{2} + t^{2}\right)\right\}}
\end{equation}
for the ground-state wave function.  What do Feynman {\it et al.}
say about this oscillator wave function?

\begin{figure}%[thb]
\centerline{\epsfig{figure=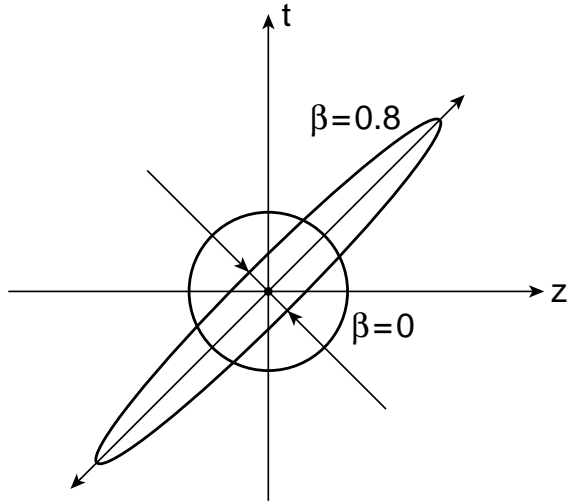,angle=0,width=80mm}}
\caption{Effect of the Lorentz boost on the space-time
wave function.  The circular space-time distribution at the rest frame
becomes Lorentz-squeezed to become an elliptic
distribution.}\label{ellipse}
\end{figure}

In their classic 1971 paper~\cite{fkr71}, Feynman {\it et al.} start
with the following Lorentz-invariant differential equation.
\begin{equation}\label{osceq}
{1\over 2} \left\{x^{2}_{\mu} -
{\partial^{2} \over \partial x_{\mu }^{2}}
\right\} \psi(x) = \lambda \psi(x) .
\end{equation}
This partial differential equation has many different solutions
depending on the choice of separable variables and boundary conditions.
Feynman {\it et al.} insist on Lorentz-invariant solutions which are
not normalizable.  On the other hand, if we insist on normalization,
the ground-state wave function takes the form of Eq.(\ref{ground}).
It is then possible to construct a representation of the
Poincar\'e group from the solutions of the above differential
equation~\cite{knp86}.  If the system is boosted, the wave function
becomes
\begin{equation}\label{eta}
\psi_{\eta }(z,t) = \left({1 \over \pi }\right)^{1/2}
\exp\left\{-{1\over 2}\left(e^{-2\eta }u^{2} +
e^{2\eta}v^{2}\right)\right\} .
\end{equation}
This wave function becomes Eq.(\ref{ground}) if $\eta$ becomes zero.
The transition from Eq.(\ref{ground}) to Eq.(\ref{eta}) is a
squeeze transformation.  The wave function of Eq.(\ref{ground}) is
distributed within a circular region in the $u v$ plane, and thus
in the $z t$ plane.  On the other hand, the wave function of
Eq.(\ref{eta}) is distributed in an elliptic region with the light-cone
axes as the major and minor axes respectively.  If $\eta$ becomes very
large, the wave function becomes concentrated along one of the
light-cone axes.  Indeed, the form given in Eq.(\ref{eta}) is a
Lorentz-squeezed wave  function.  This squeeze mechanism is
illustrated in Fig.~\ref{ellipse}.

There are many different solutions of the Lorentz invariant differential
equation of Eq.(\ref{osceq}).  The solution given in Eq.(\ref{eta})
is not Lorentz invariant but is covariant.  It is normalizable
in the $t$ variable, as well as in the space-separation variable $z$.
How can we extract probability interpretation from this covariant
wave function?

%----------------------------------------------------------------------

%----------------------------------------------------------------------

\section{Probability Interpretation}\label{prob}
The wave function of Eq.(\ref{eta}) is a Lorenntz-covariant expression.
To an observer in which the hadron is at rest, the parameter $\eta$
becomes zero.  For an observer in which the hadron moves very
fast, $\eta$ becomes very large.  The question then is whether
it is possible to construct a probability distribution function
from this wave function which contains the time-separation variable
$t$.

At this point, it is very important to note that the step toward the
construction of the probability distribution not always to take the
absolute square of the wave function.  The first step is to construct
the density matrix which in turn gives the probability distribution.
The second step is to sum over or integrate over the variables which
are not observed.

Here again let us learn a lesson from Feynman.  In his 1972 book on
statistical mechanics~\cite{fey72}, Feynman says {\it When we solve
a quantum-mechanical problem, what
we really do is divide the universe into two parts -- the system
in which we are interested and the rest of the universe.  We then
usually act as if the system in which we are interested comprised
entire universe.  To motivate the use of density matrices, let us
see what happens when we include the part of the universe outside
the system.}

It is an interesting exercise to work out a concrete example of
Feynman's rest of the universe.  Let us consider two identical
coupled oscillators with coordinates $x_{1}$ and $x_{2}$.
Here the first coordinate is in the system in which we are interested
and the second belongs to the rest of the universe.  In order to
get the probability distribution in the first coordinate while
it is not possible to get any information from the second coordinate,
we can construct the density matrix first with both coordinate
variables and then integrate over the second variable.  The effect
of this ignorance is measured in terms of the entropy.  This procedure
has been discussed in detail in Ref.~\cite{hkn99ajp}.  Let us outline
the content of this paper.

For the coupled harmonic oscillators, the standard procedure is to
separate the Hamiltonian using the normal coordinates
$\left(x_{1} + x_{2}\right)/\sqrt{2}$ and
$\left(x_{1} - x_{2}\right)/\sqrt{2}$.  Then the resulting
wave function is
\begin{equation}\label{eta2}
\psi_{\eta }\left(x_{1}, x_{2}\right) =
\left({1 \over \pi }\right)^{1/2}
\exp\left\{-{1\over 4}\left(e^{-2\eta }
\left(x_{1} + x_{2}\right)^{2} +
e^{2\eta}  \left(x_{1} - x_{2}\right)^{2}\right)\right\} ,
\end{equation}
where the parameter $\eta$ in this case measures the strength of
coupling.  If it is zero, the oscillator system becomes decoupled.
Here, it is possible to give probability interpretation in terms
of the two normal coordinates as well as in terms of the
$x_{1}$ and $x_{1}$ variables.

It is indeed remarkable that the wave function of Eq.(\ref{eta2})
is mathematically identical with the covariant oscillator wave
function given in Eq.(\ref{eta}).  Thus, for the case of
the covariant oscillator, we can follow the same logic as in the
coupled oscillators.  We can now change the variable $x_{1}$ to $z$,
the variable in the system we are interested, and $x_{2}$ to $t$
which belongs to the rest of the universe.

With these ingredients in mind, let us start with the square of the
covariant wave function of Eq.(\ref{eta}):
\begin{equation}\label{rho}
\rho_{\eta}(z,t) = \psi_{\eta}(z,t) \left(\psi_{\eta}(z,t)\right)^*
\end{equation}
This is the density matrix if both the $z$ and $t$ variables are
taken into account.  Since we do not observe $t$ in the present form
of quantum mechanics, we have to integrate the above quantity over
the $t$ variable.
\begin{equation}\label{rho2}
\rho_{\eta}(z) = \int \rho_{\eta}(z,t) dt .
\end{equation}
The evaluation of this integral leads to the probability density:
\begin{equation}
\rho_{eta}(z) = \left({1 \over \pi (\cosh2\eta)} \right)^{1/2}
   \exp{\left( {- z^{2} \over \cosh2\eta} \right)} .
\end{equation}
This is a normalized probability distribution function.  If the
hadronic speed is zero, this expression reduces to the probability
density for the ground-state harmonic oscillator.

The wave function becomes wide-spread as $\eta$ increases, but the
normalization is preserved.  The Lorentz boost is therefore a
probability-preserving transformation.

In Ref.~\cite{hkn99ajp} in which the coupled oscillator system is
discussed, it was observed that our ignorance over the rest of the
universe causes an increase in entropy.  The expression
of the entropy does not contain the coordinate variables there.
Thus it should also be independent of $z$ and $t$ variables.
The entropy for this system is
\begin{eqnarray}
S = -Tr(\rho~\ln(\rho)) \nonumber \\[2ex]
= - \int \rho (z) \ln \left(\rho(z)\right) dz .
\end{eqnarray}
The evaluation of integral leads to
\begin{equation}
S = \left(\cosh{\eta \over 2}\right)^{2}
\ln\left( \cosh{\eta \over 2}\right)^{2} -
\left( \sinh{\eta \over 2}\right)^{2}
\ln\left( \sinh{\eta \over 2}\right)^{2} .
\end{equation}
This form is the same as the entropy for thermally-excited
oscillator state.  The entropy vanishes when the hadron is at rest
with $\eta = 0$, and increases linearly in $\eta$ as the hadronic
speed becomes large.

We learned in this section how to take care of the time-separation
variable when we obtain the probability distribution from the
covariant wave function of Eq.(\ref{eta}).   We learned also that
it is now possible to give a probability interpretation of the
density function of Eq.(\ref{rho}) in terms of both the $z$ and
$t$ variables, as in the case of $x_{1}$ and $x_{2}$ variables for
the coupled oscillator case.

\section{Feynman's Parton Picture}\label{par}

It is a widely accepted view that hadrons are quantum bound states
of quarks having localized probability distribution.  As in all
bound-state cases, this localization condition is responsible for
the existence of discrete mass spectra.  The most convincing evidence
for this bound-state picture is the hadronic mass spectra which are
observed in high-energy laboratories~\cite{fkr71,knp86}.
However, this picture of bound states is applicable only to observers
in the Lorentz frame in which the hadron is at rest.  How would the
hadrons appear to observers in other Lorentz frames?  To answer this
question, we can use the picture of Lorentz-squeezed hadrons discussed
in Sec.~\ref{prob}.

In 1969, Feynman observed that a fast-moving hadron can be regarded
as a collection of many ``partons'' whose properties not appear to be
quite different from those of the quarks~\cite{fey69}.  For example,
the number of quarks inside a static proton is three, while the number
of partons in a rapidly moving proton appears to be infinite.  The
question then is how the proton looking like a bound state of quarks
to one observer can appear different to an observer in a different
Lorentz frame?  Feynman made the following systematic observations.

\begin{itemize}

\item[a.]  The picture is valid only for hadrons moving with
  velocity close to that of light.

\item[b.]  The interaction time between the quarks becomes dilated,
   and partons behave as free independent particles.

\item[c.]  The momentum distribution of partons becomes widespread as
   the hadron moves fast.

\item[d.]  The number of partons seems to be infinite or much larger
    than that of quarks.

\end{itemize}

\noindent Because the hadron is believed to be a bound state of two
or three quarks, each of the above phenomena appears as a paradox,
particularly b) and c) together.

In order to resolve this paradox, let us write down the
momentum-energy wave function corresponding to Eq.(\ref{eta}).
If the quarks have the four-momenta $p_{a}$ and $p_{b}$, we can
construct two independent four-momentum variables~\cite{fkr71}
\begin{equation}
P = p_{a} + p_{b} , \qquad q = \sqrt{2}(p_{a} - p_{b}) ,
\end{equation}
where $P$ is the total four-momentum and is thus the hadronic
four-momentum.  $q$ measures the four-momentum separation between
the quarks.  Their light-cone variables are
\begin{equation}\label{conju}
q_{u} = (q_{0} - q_{z})/\sqrt{2} ,  \qquad
q_{v} = (q_{0} + q_{z})/\sqrt{2} .
\end{equation}
The resulting momentum-energy wave function is
\begin{equation}\label{phi}
\phi_{\eta }(q_{z},q_{0}) = \left({1 \over \pi }\right)^{1/2}
\exp\left\{-{1\over 2}\left(e^{-2\eta}q_{u}^{2} +
e^{2\eta}q_{v}^{2}\right)\right\} .
\end{equation}
Because we are using here the harmonic oscillator, the mathematical
form of the above momentum-energy wave function is identical to that
of the space-time wave function.  The Lorentz squeeze properties of
these wave functions are also the same.  This aspect of the squeeze
has been exhaustively discussed in the
literature~\cite{knp86,kn77par,kim89}.

When the hadron is at rest with $\eta = 0$, both wave functions
behave like those for the static bound state of quarks.  As $\eta$
increases, the wave functions become continuously squeezed until
they become concentrated along their respective positive
light-cone axes.  Let us look at the z-axis projection of the
space-time wave function.  Indeed, the width of the quark distribution
increases as the hadronic speed approaches that of the speed of
light.  The position of each quark appears widespread to the observer
in the laboratory frame, and the quarks appear like free particles.

%----------------------------------------------------------------------
\begin{figure}%[thb]
\centerline{\epsfig{figure=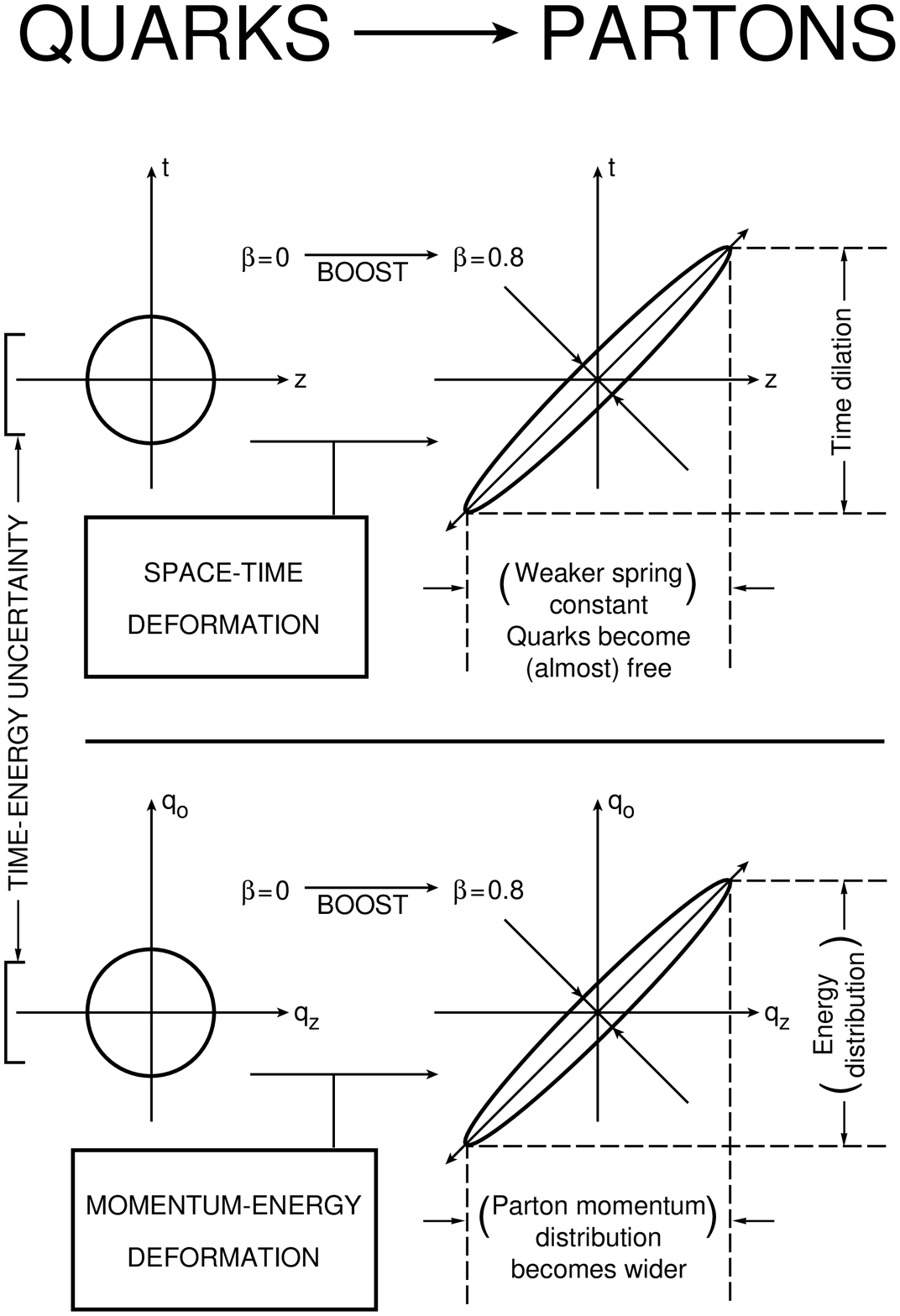,angle=0,width=90mm}}
\vspace{5mm}
\caption{Lorentz-squeezed space-time and momentum-energy wave
functions.  As the hadron's speed approaches that of light, both
wave functions become concentrated along their respective positive
light-cone axes.  These light-cone concentrations lead to Feynman's
parton picture.}\label{parton}
\end{figure}
%----------------------------------------------------------------------

The momentum-energy wave function is just like the space-time wave
function, as is shown in Fig.~\ref{parton}.  The longitudinal momentum
distribution becomes wide-spread as the hadronic speed approaches the
velocity of light.  This is in contradiction with our expectation from
non-relativistic quantum mechanics that the width of the momentum
distribution is inversely proportional to that of the position wave
function.  Our expectation is that if the quarks are free, they must
have their sharply defined momenta, not a wide-spread distribution.

However, according to our Lorentz-squeezed space-time and
momentum-energy wave functions, the space-time width and the
momentum-energy width increase in the same direction as the hadron
is boosted.  This is of course an effect of Lorentz covariance.
This indeed is the key to the resolution of the quark-parton
paradox~\cite{knp86,kn77par}.

\section{Measurement Problem}\label{measure}

According to Sec.~\ref{prob}, it is possible to define the covariant
probability distribution function in terms of both the space and
time separation variables.  Indeed, the quark distribution can be
covariantly transformed according to Fig.~\ref{parton}.  In this
section, we address the question of how partons interact incoherently
with external signals.

When the hadron is boosted, the hadronic matter becomes squeezed and
becomes concentrated in the elliptic region along the positive
light-cone axis, as is illustrated in Figs.~\ref{ellipse} and
\ref{parton}.  The length of the major axis becomes expanded by
$e^{\eta}$, and the minor axis is contracted by $e^{-\eta}$.

This means that the interaction time of the quarks among themselves
becomes dilated.  Because the wave function becomes wide-spread, the
distance between one end of the harmonic oscillator well and the
other end increases.  This effect, first noted by Feynman~\cite{fey69},
is universally observed in high-energy hadronic experiments.  The
period of the oscillation increases like $e^{\eta}$.

On the other hand, the interaction time with the external signal,
since it is moving in the direction opposite to the direction of
the hadron, travels along the negative light-cone axis.
If the hadron contracts along the negative light-cone axis, the
interaction time decreases by $e^{-\eta}$.  The ratio of the interaction
time to the oscillator period becomes $e^{-2\eta}$.  The energy of each
proton coming out of the Fermilab accelerator is $900 GeV$.  This leads
the ratio to $10^{-6}$.  This is indeed a small number.  The external
signal is not able to sense the interaction of the quarks among
themselves inside the hadron.

Indeed, Feynman's parton picture is one concrete physical example
where the decoherence effect is observed.  As for the entropy, the
time-separation variable belongs to the rest of the universe.  Because
we are not able to observe this variable, the entropy increases
as the hadron is boosted to exhibit the parton effect.  The
decoherence is thus accompanied by an entropy increase.

Let us go back to the coupled-oscillator system.  The light-cone
variables in Eq.(\ref{eta}) correspond to the normal coordinates in
the coupled-oscillator system given in Eq.(\ref{normal}).  According
to Feynman's parton picture, the decoherence mechanism is determined
by the ratio of widths of the wave function along the two normal
coordinates, or along the two light-cone coordinates.  Indeed,
Feynman's decoherence is observed in the light-cone coordinate
system.

\section*{Concluding Remarks}

The present authors have been interested in question of covariant
probability distribution since 1973~\cite{kn73}.  We started
with the covariant oscillator wave function as a purely
phenomenological mathematical instrument.  We then noticed that
the covariant oscillator formalism can serve as a representation
of the Wigner's little group for massive particles, capable of
the fundamental symmetry representation for relativistic particles.
This allows us to deal with the c-number time-energy uncertainty
relation without excitations.

Later, we were able to illustrate Feynman's rest of the universe
in terms of two coupled oscillators.  By comparing the covariant
oscillator with the coupled oscillator, we are able to give
a covariant probability interpretation to the density matrix
depending on both the space-separation and time-separation
variables.

This covariant probability distribution function can provide a
resolution to the quark-parton puzzle which includes Feynman's
decoherence.

\end{document}